\documentclass[trackchanges]{aastex701}
\usepackage{amsmath}
\usepackage{float}

\begin{document}

\title{ Determining the Spin and Orbital Period of the Asynchronous Cataclysmic Variable X10 in 47 Tucanae}

\author[]{Daniel Mendoza-Fredes}
\affiliation{Physics Dept., University of Alberta,
CCIS 4-181, Edmonton, Alberta, Canada, T6G 2E1}
\affiliation{Departamento de Astronom\'{i}a, Universidad de Concepci\'{o}n, Av. Esteban Iturra s/n Barrio Universitario, Casilla 160-C, Concepci\'{o}n, Chile}
\email[show]{damendoza2021@udec.cl}  

\author[orcid=0000-0003-3944-6109]{Craig O. Heinke} 
\affiliation{Physics Dept., University of Alberta,
CCIS 4-181, Edmonton, Alberta, Canada, T6G 2E1}
\email{heinke@ualberta.ca}

\author[]{Matthew Lang}
\affiliation{Physics Dept., University of Alberta,
CCIS 4-181, Edmonton, Alberta, Canada, T6G 2E1}
\email{mflang@ualberta.ca}

\author[]{Asma Hattawi}
\affiliation{Sharjah Space and Astronomy Hub  (SSAH), University of Sharjah, Sharjah City, Sharjah, 27272, UAE}
\affiliation{Physics Dept., University of Alberta,
CCIS 4-181, Edmonton, Alberta, Canada, T6G 2E1}
\email{ahattawi@sharjah.ac.ae}

\begin{abstract}

The X-ray source X10 in the globular cluster 47 Tuc has been identified as a cataclysmic variable (CV). With a suggested 4.7 hour periodicity, and a soft X-ray spectrum, it appears to be an unusually X-ray luminous ($L_X=2.2\times10^{32}$ erg s$^{-1}$) polar CV.
We use 286 ks of Chandra X-ray Observatory observations  of 47 Tuc to study the X-ray timing behavior of 47 Tuc X10. 
X10 shows clear pulsations in the data, with occasional phase shifts.  These suggest that X10 is an asynchronous polar CV, where the spin and orbital periods are slightly misaligned, and the dominant accretion flow shifts from one pole to the other. 

We use a Lomb-Scargle periodogram to suggest potential spin and orbital periods, then model the X-ray lightcurve using simple sinusoids to represent emission from each accretion pole, switching which pole is visible on the beat period. We find a white dwarf spin period of $P_{\mathrm{spin}} = 4.2683$ h and derive an orbital period of $P_{\mathrm{orb}} =  4.6779$ h from the beat modulation.
This large degree of asynchronism (8.8~\%) suggests that X10 is in an unstable equilibrium, 
and may be a transitional system between an intermediate polar and a fully synchronized polar.


\end{abstract}

\keywords{\uat{Cataclysmic variable stars}{203} --- \uat{AM Herculis stars}{32} --- \uat{X-ray astronomy}{1810} --- \uat{Globular star clusters}{656} --- \uat{Period determination}{1211} }


\section{Introduction} 

Cataclysmic Variables (CVs) are binary systems consisting of a white dwarf (WD) and a low-mass late-type secondary star, which fills its Roche lobe, allowing mass transfer onto the WD \citep{Warner95}. If the WD has a relatively weak magnetic field, the accretion flow forms a disk that reaches down to the surface of the WD, producing radiation that typically peaks in the UV. If the mass transfer rate is not high enough to keep material in the disk ionized, material will build up in the disk until the temperature and density cause ionization, and an increase in viscosity, which causes the disk to empty onto the WD. This is referred to as a dwarf nova outburst \citep{Lasota01}.

 In systems where the WD has a very strong magnetic field ($B \gtrsim 10$ MG), accretion disks do not form, as the magnetic field disrupts the flow when magnetic forces exceed the ram pressure of the infalling material, forcing the accreting matter to follow magnetic field lines and accrete at the magnetic poles. 
 These systems are known as polars, or AM Herculis stars. In a standard polar, the magnetic interaction between the two stars is sufficient to lock the WD's rotation to the binary orbit, resulting in synchronous rotation ($P_{\mathrm{spin}} = P_{\mathrm{orb}}$) \citep{cropper90}. Most polars have periods of order 90-200 minutes, and relatively low accretion rates, with typical $L_X$ within $10^{30}-10^{32}$  erg s$^{-1}$ \citep{Schwope24}. 
 This class is distinct from the ``intermediate polars", which also have (typically slightly less strong) magnetic fields that channel the accretion flow to the WD poles, but do not lock the WD spin and orbital period \citep{Patterson94}. Intermediate polars typically have short spin periods of 1-30 minutes, orbital periods of $>3$ hours (and  higher accretion rates than the polars), and tend to be close to spin equilibrium \citep{Patterson2020}. A large fraction of known (often X-ray-selected) intermediate polars have relatively high $L_X$ of $10^{33}-10^{34}$  erg s$^{-1}$ \citep{Suleimanov19}. 


However, a subclass of polars known as asynchronous polars (APs) exhibits a slight difference between the white dwarf spin and orbital period \citep{Campbell99}. The majority of these APs show a difference between their spin ($\omega$) and orbital ($\Omega$) periods of $<$2\%.  This state is often attributed to recent nova eruptions that perturb the white dwarf's rotation, though the systems eventually evolve back toward synchronism over thousands of years \citep{Myers_2017}.  A few other systems show larger (10-20\%) differences between their spin and orbital periods, which have been harder to explain \citep{Schwarz07,Halpern22,Littlefield23a}. 
Because the white dwarf rotates relative to the secondary star, the orientation of its magnetic axis with respect to the companion star (and accretion flow) changes, which can allow the accretion stream to periodically switch between opposite magnetic poles, a phenomenon known as pole-switching \citep{halpern_resolving_2024}. This process results in complex X-ray and optical light curves modulated by the beat frequency $\omega - \Omega$ \citep{wynn, Wang_2020}. In most cases these periodicities are unveiled by identifying peaks in periodograms from long optical time series \citep[e.g.][]{Schwarz07,Myers_2017,Tovmassian17,Littlefield23a,Littlefield23b}. However, in some cases evidence for pole-switching has come from  inspection and periodograms of X-ray lightcurves \citep{Bernardini13,Joshi16,Halpern17,Rea17}. 

A large number of CVs have been discovered in Galactic globular clusters \citep[e.g.][]{Knigge2012}, the majority by initial detection in X-rays with the Chandra X-ray Observatory and identification with blue, variable stars in Hubble Space Telescope (HST) imaging \citep{Grindlay_2001,Pooley02}.  The globular cluster 47 Tuc has particularly deep X-ray and optical/UV datasets, as a dense, massive cluster only 4.45 kpc from us \citep{Harris10,Chen18}. The ROSAT X-ray survey of 47 Tuc identified nine X-ray sources in and around the core of 47 Tuc \citep{Hasinger94,Verbunt98}. One of these sources, X10, matches the faint blue variable  V3 identified with HST data as a likely CV \citep{Shara96}. Additional HST imaging confirmed a 4.7-hour photometric period \citep{Edmonds03a,Edmonds03b}, blue UV colors and H-$\alpha$ emission \citep{RiveraSandoval18}. 
Chandra imaging confirmed the identification, and revealed that X10 (aka CXOGlb J002406.3-720443, or W27) showed X-ray pulsations with a 4.7 hour period, and a very soft X-ray spectrum, indicating that it is a polar CV \citep{Heinke2005}.
The X-ray spectra from the 2002 Chandra data showed a luminous blackbody component with $kT=53$ eV, plus two optically thin thermal plasma components with temperatures 0.39$\pm0.03$ and $>$14 keV \citep{Heinke2005}. 
X10 is unusually luminous for a polar, or for a globular cluster CV, at $L_X$(0.5-6 keV)=$2.2\times10^{32}$  erg s$^{-1}$ \citep{Heinke2005}. However, the long Chandra observations revealed that X10's pulsations showed evidence of pole-switching, suggesting that more careful fitting was needed to infer the true spin and orbital periods \citep{hattawi2019x10}.

Recently several additional probable polar CVs have been suggested in 47 Tuc, using X-ray timing; 47 Tuc W2 \citep{Bao23,Amato24}, and W53, W20, W23 \citep{Bao23}.   \citet{Bao24} identified, via periodic signals, five more likely polar CVs in other globular clusters, and two candidate intermediate polars. 

Whether globular cluster CVs are more, or less, likely to be magnetic than CVs elsewhere in the galaxy has long been a subject for debate. \citet{Shara96} found only one dwarf nova eruption in 47 Tucanae, arguing that there were far fewer non-magnetic CVs than expected from theories of dynamical formation of CVs in globular clusters. \citet{Grindlay95} and \citet{Edmonds99} used HST spectra of 4 CV candidates in NGC 6397, showing He II 4696 emission lines, to argue that these CVs were magnetic, since the high temperatures required to produce these lines can be produced at the accreting magnetic poles. 

\citet{Edmonds03b} noticed that HST CV candidate counterparts to Chandra X-ray sources in 47 Tuc were on average optically fainter than known CVs (including intermediate polars) outside clusters, but had higher X-ray/optical flux ratios than other CVs. \citet{Dobrotka06} presented simulations arguing that some degree of magnetism is necessary for globular cluster CVs to show relatively few dwarf nova outbursts. \citet{Ivanova06} suggested that the higher X-ray/optical ratio in globular cluster CVs may be due to higher average masses in WDs that are exchanged via dynamical interactions into CVs, and that since higher-mass WDs are more likely magnetic, that globular CVs are more likely to be magnetic. \citet{Heinke08} argued from the X-ray spectra of globular cluster CVs that perhaps $\sim$40\% of this (X-ray selected) sample are magnetic. However, \citet{Bahramian2020} showed that globular clusters are remarkably underabundant in X-ray luminous ($L_X>10^{33}$ erg s$^{-1}$) intermediate polars, compared to local space. \citet{Belloni21} explained this paucity of bright intermediate polars as due to the full crystallization of WDs before mass transfer begins in dynamically-formed CV systems.    
It is not yet clear whether the formation of magnetic CVs is enhanced or not in clusters, or how magnetic CVs in clusters may be different from magnetic CVs elsewhere.


This paper attempts to infer the spin and orbital periods of X10, using the long series of Chandra observations in 2002. Using a 20-parameter sinusoidal model and the \texttt{emcee} Markov Chain Monte Carlo (MCMC) sampler \citep{Foreman_Mackey_2013}, we explore the parameter space to resolve the ambiguity between the true physical periods and their associated sidebands.


\section{Observational Data}

\begin{figure}[ht!]
    \centering
    \includegraphics[width=\textwidth]{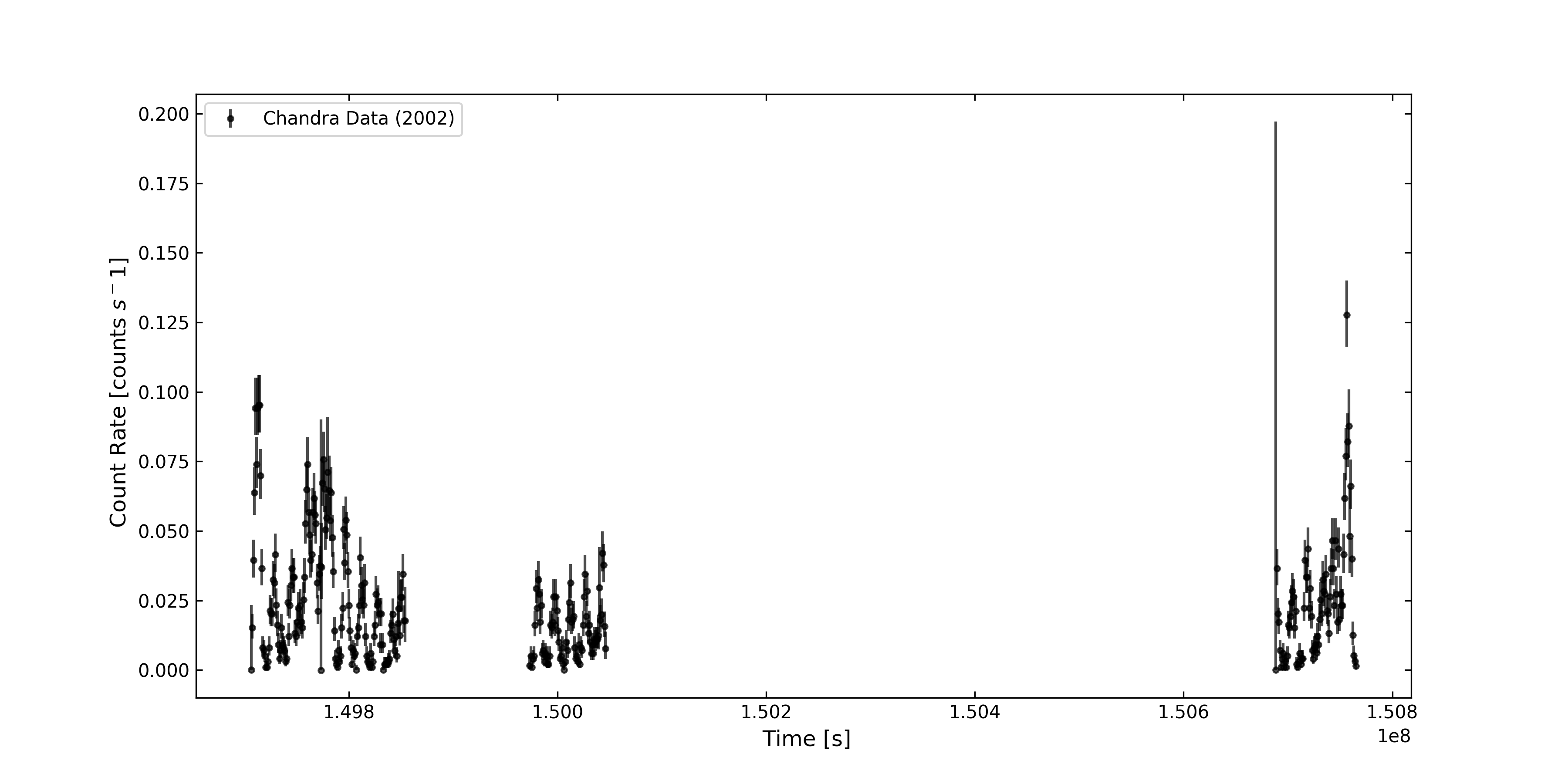}
    \caption{Merged Chandra ACIS-S 0.3-8 keV X-ray light curve of X10, from the 2002 epoch. The dataset combines the eight distinct observations. Note the significant time gaps between observing windows.}
    \label{fig:raw_2002}
\end{figure}

Multiple high-resolution X-ray imaging surveys of 47 Tucanae (NGC 104) have been performed using the ACIS detector \citep{garmire2003} of the Chandra X-ray Observatory \citep{Weisskopf02}. 
We focused our analysis on the 2002 dataset (ObsIDs 2735-2738 and 3384-3387), presented in \citet{Heinke2005}; see their Table 1. 
This data consists of 4 70-ks observations spaced logarithmically in time, each followed by 5-ks shorter observations using a detector subarray. The first two observation pairs create nearly 150 ks of essentially uninterrupted observation.  
The 2002 data were selected because they are the first, and longest, series of observations using the most sensitive ACIS-S detector on Chandra. 
This dataset shows the most well-defined sinusoidal pulsations, and evidence for pole-switching, allowing for more robust modeling compared to other epochs. 
Fig.\ \ref{fig:raw_2002} shows a lightcurve of the 2002 data as a whole.
This paper employs a list of Chandra datasets, obtained by the Chandra X-ray Observatory, contained in the Chandra Data Collection (CDC) 724~\dataset[doi:10.25574/cdc.724]{https://doi.org/10.25574/cdc.724}.

\section{Methods}

\subsection{Data preparation and Error Analysis}

We used version 3.3 of the CIAO\footnote{\url{https://cxc.cfa.harvard.edu/ciao/releasenotes/ciao_3.3_release.html}} software \citep{Fruscione26} to barycenter each observation (using the {\tt axbary} command), and extract event files and lightcurves,  using a 1" extraction region. 
We created lightcurves between 0.1 and 7 keV, with 1000 second resolution. 
Upper and lower error bars on the counts per bin were computed according to the definition given by \cite{Gehrels86} for small number of events based on Poisson statistics;

\begin{equation}\label{upper}
\lambda_u \approx n + S \sqrt{\,n + \frac{3}{4}} + \frac{S^2 + 3}{4} \, .
\end{equation}

\begin{equation}\label{lower}
\lambda_\ell \approx n - S \sqrt{\,n - \frac{1}{4}} + \frac{S^2 - 1}{4} \, .
\end{equation}

\noindent where Equation \ref{upper} represents the single-sided upper limit ($\lambda_u$) and Equation \ref{lower} defines the single-sided lower limit ($\lambda_\ell$). In these expressions, $n$ represents the number of detected events (counts), and $S$ is the significance level in Gaussian standard deviations. For this analysis, we adopted $S=1$, corresponding to a confidence level of 84.13\%, consistent with the standard $1\sigma$ error convention in X-ray astronomy.

\subsection{Periodicity Search}

We first performed a 
power spectrum 
analysis to identify the fundamental periodicities of the system and constrain the parameter space for the MCMC simulation. We utilized the Lomb-Scargle periodogram \citep{lomb,scargle}, which is optimized for detecting periodic signals in unevenly sampled time series.

\begin{figure}[ht! ]
    \centering
    \includegraphics[width=\textwidth]{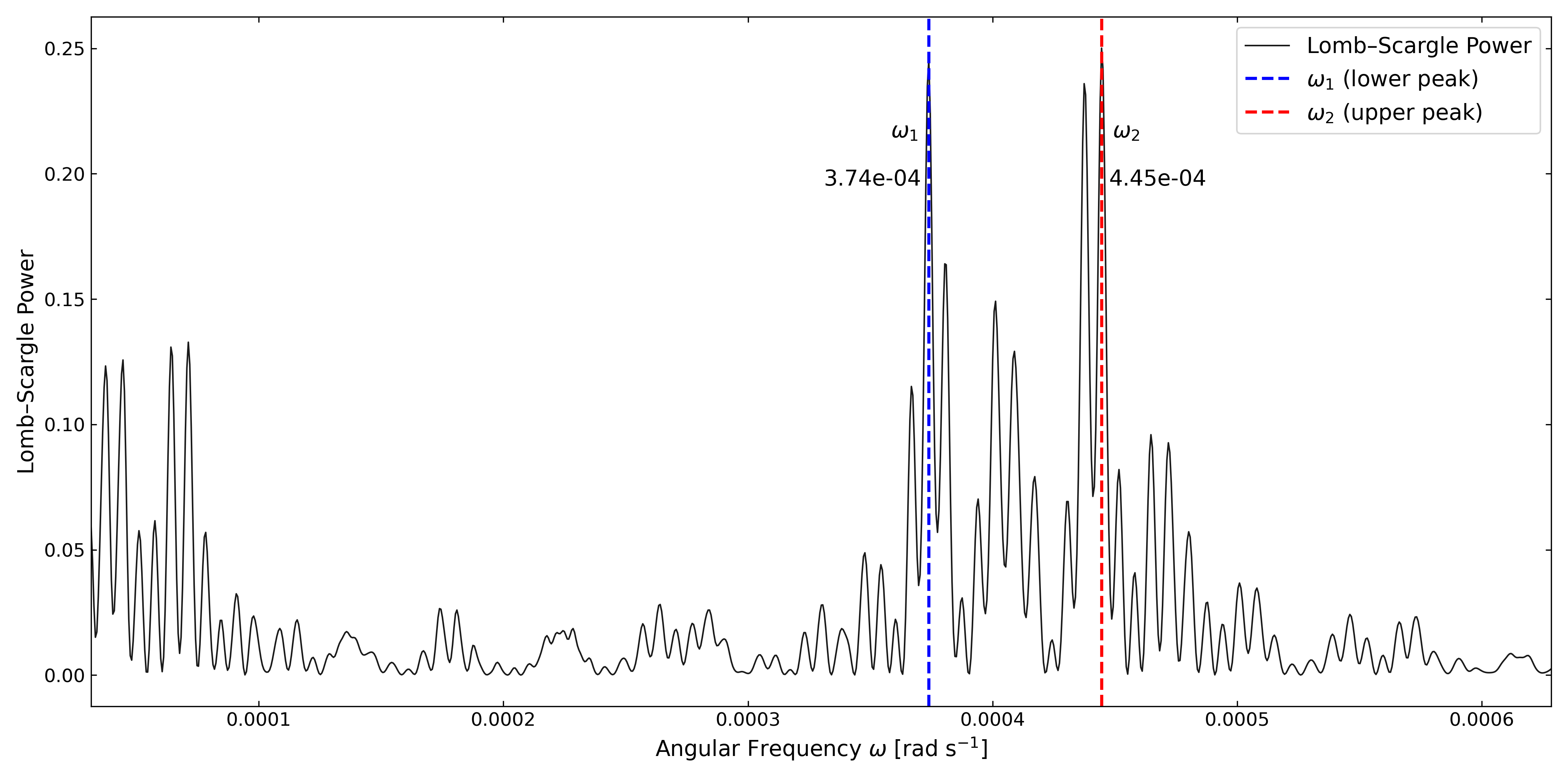}
    \caption{Lomb-Scargle periodogram of the 2002 X-ray dataset in angular frequency space. The two dominant peaks at $\omega_1$ and $\omega_2$ are indicated. }
    \label{fig:periodogram}
\end{figure}

We computed the periodogram for the merged 2002 dataset, over a frequency range consistent with typical CVs (Figure~\ref{fig:periodogram}). 
The two most prominent peaks were seen at $\omega_1=3.74\times10^{-4}$ rad s$^{-1}$ and $\omega_2=4.45\times10^{-4}$ rad s$^{-1}$, with a number of other peaks approaching their strength. Their interpretation is ambiguous from the periodogram alone. 

\subsection{The Numerical Model}

The X-ray light curve of X10 is primarily driven by the changing visibility of the two poles, due to the rotation of the WD. We model the emission from the two accretion poles using simple sinusoidal functions. The base modulation for the primary (1) and secondary (2) poles is defined as:

\begin{equation}
S_1(t) = \sin(\omega t + \phi)
\end{equation}

\begin{equation}
S_2(t) = \sin(\omega t + \phi + C_2 + \pi)
\end{equation}

\noindent where $\omega$ is the WD spin angular frequency ($2\pi/P_{\rm spin}$), $\phi$ is a reference phase, and $C_2$ represents a phase offset accounting for possible asymmetries between the locations of the two accretion poles.

The emission from each pole is not continuous, as the accretion geometry changes over the beat cycle of the system, causing accretion onto each pole to switch on and off. To account for this behaviour, we introduce two visibility functions, $B_1(t)$ and $B_2(t)$, which modulate the contribution of each pole as a function of the beat phase:

\begin{equation}
B_1(t)=1+{\rm floor}\left[\frac{1}{2}\sin\!\left(\frac{2\pi t}{P_{\mathrm{beat}}}+H_3\right)+H_1 \right]
\end{equation}

\begin{equation}
B_2(t)=1+{\rm floor}\left[ \frac{1}{2}\sin\!\left(\frac{2\pi t}{P_{\mathrm{beat}}}+\pi+H_4\right)+H_2 \right]
\end{equation}

\noindent where $P_{\rm beat}$ is the beat period, $H_1$ and $H_2$ control the duration of visibility of each pole, and $H_3$ and $H_4$ set the phase alignment of the pole-switching behaviour. Each function takes a value of 1 when the sinusoid is positive and 0 otherwise, thereby regulating visibility of each pole. 

To account for the background level when the poles are not visible or during minima in the pulse profile, we introduce a constant background, $E$, adopting a fixed value of $0.00227$ counts s$^{-1}$. This baseline corresponds to the mean flux of all data points during quiescent intervals, defined by a threshold below $ 0.005$ counts s$^{-1}$. At each time step, the model evaluates the total emission by selecting the maximum value between the modulated flux and this background level $E$.

Finally, since the X-ray intensity of X10 varies significantly between observations, a single constant amplitude is insufficient to reproduce the long-term variability. We therefore model the amplitude of the poles, $M_i(t)$, as a time-dependent piecewise function that allows for linear evolution within individual observing windows:

\begin{equation}\label{piecewise_amp}
M_i(t) =
\begin{cases}
A_3 + G_3 \cdot t & \text{for } 0 \le t < 50\,000 \text{ s} \\
A_4 + G_4 \cdot t & \text{for } 50\,000 \le t < 83\,000 \text{ s} \\
A_5 + G_5 \cdot t & \text{for } 83\,000 \le t < 160\,000 \text{ s} \\
A_6 + G_6 \cdot t & \text{for } 260\,000 \le t < 350\,000 \text{ s} \\
A_{1,2} + G_{1,2} \cdot (t - t_{\mathrm{ref}}) & \text{for } t \ge 9.8\times10^{5} \text{ s}
\end{cases}
\end{equation}

\noindent where $i = 1, 2$ labels the accretion pole,  $A_1$ represents the base amplitude within a specific interval, and $G_i$ gives the slope of  fading or brightening during that time. $t$ is measured from the start of the 2002 observations.  
The intervals defined in Equation~\ref{piecewise_amp} cover all but the last long 2002 observation. 
In the final observation, the two poles are allowed to evolve independently, with slopes $G_1$ and $G_2$. The reference epoch $t_{\mathrm{ref}} = 1.02 \times 10^{6}$ s is the midpoint of that observation, so that $A_1$ and $A_2$ correspond to the pole amplitudes at $t_{\mathrm{ref}}$ rather than an extrapolated value. Amplitudes are constrained 
to be non-negative, 
so that their evolution cannot produce negative values.

Combining the intrinsic spin sinusoid ($S$), the visibility switches ($B$), and the background ($E$), the final model for the count rate $R(t)$ is given by:

\begin{equation}\label{model}
    R(t) = \max\left[M_1(t) B_1(t) S_1(t), 0\right] + \max\left[M_2(t) B_2(t) S_2(t), 0\right] +E
\end{equation}

\noindent where $M_1(t)$ and $M_2(t)$ represent the time-dependent amplitudes of each accretion pole.

\subsubsection{Beat Period}

The beat period ($P_{ beat}$) represents the timescale over which the relative phase between the white dwarf spin and the orbital motion of the companion star repeats. In frequency space, it corresponds to the difference between the spin angular frequency ($\omega$) and the orbital angular frequency ($\Omega$):

\begin{equation}\label{fred}
    \omega_{ beat} = \omega - \Omega
\end{equation}

\noindent Since angular frequency can be expressed in terms of period as $\omega = 2\pi/P$, the beat period is given by:

\begin{equation}\label{beat_period}
    \frac{1}{P_\mathrm{ beat}} = \frac{1}{P_\mathrm{spin}} - \frac{1}{P_{\mathrm{orb}}}
\end{equation}

\noindent where we assume $P_{\mathrm{spin}} < P_{\mathrm{orb}}$. These relations are central to our analysis.  The MCMC model directly fits the spin period $P_{\mathrm{spin}}$, which governs the intrinsic modulation, and the beat period $P_{ beat}$, which controls the pole-switching visibility. The orbital period $P_{\mathrm{orb}}$ is typically not directly detectable in the power spectrum, and is therefore derived indirectly using Equations \ref{fred} or \ref{beat_period}.





\subsection{MCMC Configuration}

To constrain the model parameters and estimate their uncertainties, we employed the Affine Invariant Markov Chain Monte Carlo (MCMC) ensemble sampler implemented in the Python package \texttt{emcee} \citep{Foreman_Mackey_2013}. This approach enables an efficient exploration of the multi-dimensional parameter space defined by the numerical model and allows  estimation of parameter uncertainties based on the sampled distributions.

\subsubsection{Priors}

The parameter ranges define the physical boundaries of the search space explored by the MCMC sampler. We adopted uniform parameter ranges for all model parameters, chosen based on physical considerations and previous observational constraints. For example, the spin period range was centered around the dominant periodicities at 14,135 and 16,805 seconds identified in the Lomb-Scargle periodogram (Figure \ref{fig:periodogram}).  
The beat period was constrained by the presence of a pole switch at the end of the first 2002 observation (ObsID 2735), and the necessity of another pole switch between the 2nd and 3rd 
long exposures, 
ObsIDs 2736 and 2737 (perhaps even during ObsID 3385, around t=140,000 s).

Table \ref{priors} summarizes the parameter ranges and initial values used in the sampling.  


\begin{deluxetable}{c c c}
\tablecaption{Prior distributions and initial parameter guesses.\label{priors}}

\tablehead{
  \colhead{\textbf{Parameter}} & \colhead{\textbf{Prior Range}} & \colhead{\textbf{Initial Value}}
}

\startdata
$P_{\mathrm{spin}}$ & $\left[ 10,000,\, 20,000\right]$ s & $15,300$ s \\
$\phi$ & $[ 0, 2\pi ]$ rad & 5.14 rad\\
$P_{\mathrm{beat}}$ & $[ 140,000,\, 190,000 ]$ s & 176,250 s \\
$H_1$ & $[ 0.01, 0.5 ]$ & 0.2\\
$H_2$ &  $[ 0.01, 0.5 ]$ & 0.3\\
$H_3$ & $[ 0.1, 1.0 ]$ rad & 0.8 rad\\ 
$H_4$ & $[ 0.1, 1.5 ]$ rad & 1.0 rad\\
$C_2$ & $[ 0.001, 0.3 ]$ rad & 0.1 rad\\
$A_1$ & $[ 0.001, 0.1 ]$  $\mathrm{cts\,s^{-1}}$ & 0.03 $\mathrm{cts\,s^{-1}}$ \\
$A_2, A_5, A_6$ & $[ 0.001, 0.2 ]$ \, $\mathrm{cts\,s^{-1}}$ & 0.03 $\mathrm{cts\,s^{-1}}$ \\
$A_3$ & $[ 0.001, 0.2 ]$ \, $\mathrm{cts\,s^{-1}}$ & 0.1 $\mathrm{cts\,s^{-1}}$ \\
$A_4$ & $[ -0.1, 0.1 ]$ \, $\mathrm{cts\,s^{-1}}$ & 0.03 $\mathrm{cts\,s^{-1}}$ \\
$G_1, G_2, G_3, G_4, G_5, G_6$ & $[-5\times 10^{-6},\, 5\times 10^{-6}]$ \, $\mathrm{cts\,s^{-2}}$ & 0 \\
\enddata

\tablecomments{Prior distributions and initial parameter guesses. $H_1$ and $H_2$ are dimensionless. $H_3$, $H_4$ and $C_2$ are phase offsets in radians.}
\end{deluxetable}

The parameter space was explored using 64 walkers, running for 50,000 steps to ensure convergence.

\subsubsection{Likelihood Function}

The likelihood function $p(D|\theta)$ quantifies the probability of obtaining the observed data $D$ given a specific set of model parameters $\theta$. Since the X-ray data for X10 contain a low number of counts, we adopt a Poisson likelihood, which appropriately describes the statistics of photon-counting measurements and evaluates how well the model reproduces the observed counts \citep[e.g.][]{Cash79};

\begin{equation}\label{poisson}
    \ell(\theta) = \sum_{i=1}^{n} \left[ y_i \ln\left( f(t_i, \theta)\, E_i \right) - f(t_i, \theta)\, E_i \right]
\end{equation}

\noindent Equation~\ref{poisson} corresponds to the log-likelihood function, where $y_i$ is the number of photons (counts) detected by \textit{Chandra} in time bin $i$, $f(t_i,\theta)$ is the model count rate defined in Equation~\ref{model}, and $E_i$ is the exposure time associated with that bin. The product $f(t_i,\theta)\,E_i$ therefore represents the expected number of counts in each time interval. The term $\ln(y_i!)$, which appears in the full Poisson expression, is constant with respect to the model parameters and is omitted from the optimization process.

\subsection{Convergence}

To validate the reliability of the MCMC sampling, we first inspected the trace plots of the walkers. This visual inspection allowed us to identify the initial "burn-in" phase, where the walkers are still moving from their initial positions (Table \ref{priors}) towards the high-probability region of the parameter space. To ensure the analysis was based on the stationary posterior distribution, we discarded the first 20,000 steps as burn-in.

After the burn-in, we generated corner plots (Appendix, Fig.~\ref{fig:corner}) to visualize the posterior distributions. These allowed us to identify potential parameter degeneracies and verify that the walkers converged to a stable solution rather than going to the boundaries of the prior ranges.

Finally, we monitored standard MCMC diagnostics to assess the quality of the sampling. Visual inspection of the walker traces and posterior distributions indicates that the chains adequately explored the relevant parameter space and reached a stationary solution.

\section{Results}


\subsection{MCMC} \label{mcmcsection}


The resulting corner plot (Appendix, Figure \ref{fig:corner}) reveals the multidimensional structure of the posterior, highlighting correlations between specific parameters. The best-fit values, along with their derived $1\sigma$ uncertainties (16th and 84th percentiles), are summarized in Table \ref{tab:full_results}.

\renewcommand{\arraystretch}{1.3}
\begin{deluxetable}{c c}
\tablecaption{MCMC parameter posteriors.\label{tab:full_results}}

\tablehead{
  \colhead{\textbf{Parameter}} & \colhead{\textbf{Value}}}
\startdata
$P_{\mathrm{spin}}$ & $15,366.0^{+0.5}_{-0.6}$ s \\
$P_{\mathrm{beat}}$ & $1.7550^{+0.0026}_{-0.0037} \times 10^{5}$ s \\
$P_{\mathrm{orb}}$ & $16,840.4^{+3.5}_{-2.5}$ s \\
$\phi$ & $5.263^{+0.007}_{-0.008}$ rad \\
$H_1$ & $0.239^{+0.005}_{-0.005}$ \\
$H_2$ & $0.491^{+0.006}_{-0.010}$ \\
$H_3$ & $0.773^{+0.015}_{-0.017}$ rad \\
$H_4$ & $1.249^{+0.085}_{-0.052}$ rad \\
$C_2$ & $0.137^{+0.011}_{-0.011}$ rad \\
$A_1$ & $0.0069^{+0.0016}_{-0.0015}$ $\mathrm{cts\,s^{-1}}$ \\
$A_2$ & $0.0366^{+0.0014}_{-0.0014}$ $\mathrm{cts\,s^{-1}}$ \\
$A_3$ & $0.105^{+0.004}_{-0.004}$ $\mathrm{cts\,s^{-1}}$ \\
$A_4$ & $0.009^{+0.016}_{-0.016}$ $\mathrm{cts\,s^{-1}}$ \\
$A_5$ & $0.094^{+0.007}_{-0.007}$ $\mathrm{cts\,s^{-1}}$ \\
$A_6$ & $0.062^{+0.015}_{-0.014}$ $\mathrm{cts\,s^{-1}}$ \\
$G_1$ & $(3.20^{+0.15}_{-0.14}) \times 10^{-6}$ $\mathrm{cts\,s^{-2}}$ \\
$G_2$ & $(1.01^{+0.68}_{-0.71}) \times 10^{-7}$ $\mathrm{cts\,s^{-2}}$ \\
$G_3$ & $(-1.85^{+0.11}_{-0.11}) \times 10^{-6}$ $\mathrm{cts\,s^{-2}}$ \\
$G_4$ & $(1.04^{+0.26}_{-0.25}) \times 10^{-6}$ $\mathrm{cts\,s^{-2}}$ \\
$G_5$ & $(-5.27^{+0.58}_{-0.58}) \times 10^{-7}$ $\mathrm{cts\,s^{-2}}$ \\
$G_6$ & $(-1.22^{+0.45}_{-0.47}) \times 10^{-7}$ $\mathrm{cts\,s^{-2}}$ \\
\enddata

\tablecomments{Medians with $16\%-84\%$ confidence intervals.}
\end{deluxetable}
\renewcommand{\arraystretch}{1.0}
From the MCMC analysis, we obtained a spin period of $P_{\mathrm{spin}} = 15,366 \,\mathrm{s}$ ($\omega = 4.09 \times 10^{-4}\,\mathrm{rad\,s^{-1}}$) and a beat period of $P_{\mathrm{beat}} = 1.7550{\times10^{5}}\,\mathrm{s}$ ($\omega_{\mathrm{beat}} = 3.58 \times10^{-5}\,\mathrm{rad\,s^{-1}}$). Using the relation $\omega_{ beat} = \omega - \Omega$, we derive an orbital angular frequency of $\Omega = 3.73 \times 10^{-4}\,\mathrm{rad\,s^{-1}}$, which closely matches the lower-frequency peak $\omega_1$ observed in the periodogram (Figure \ref{fig:periodogram}).

Converting these angular frequencies into periods:

\begin{equation}
    P_{\mathrm{beat}} = 
    1.7550^{+0.0026}_{-0.0037}{\times10^{5}}
    \, \mathrm{s} 
    \quad \mathrm{or}\quad  48.751^{+0.074}_{-0.103}\,  \mathrm{h}
\end{equation}

\begin{equation}
    P_{\mathrm{spin}} = 15366.0^{+0.5}_{-0.6} \, \mathrm{s}  \quad \mathrm{or}\quad 4.2683^{+0.0001}_{-0.0002} \, \mathrm{ h} 
\end{equation}

\begin{equation}
    P_{\mathrm{orb}} = 16840.4^{+3.5}_{-2.5} \, \mathrm{s}   \quad \mathrm{or}\quad  4.6779^{+0.0010}_{-0.0007} \,\mathrm{h}
\end{equation}

Uncertainties on the derived parameters were estimated from the distribution of values sampled by the MCMC chains, using the 16th and 84th percentiles to define the quoted confidence intervals.

\subsection{Light Curve Reconstruction}

\begin{figure}
    \centering
    \includegraphics[scale=0.5]{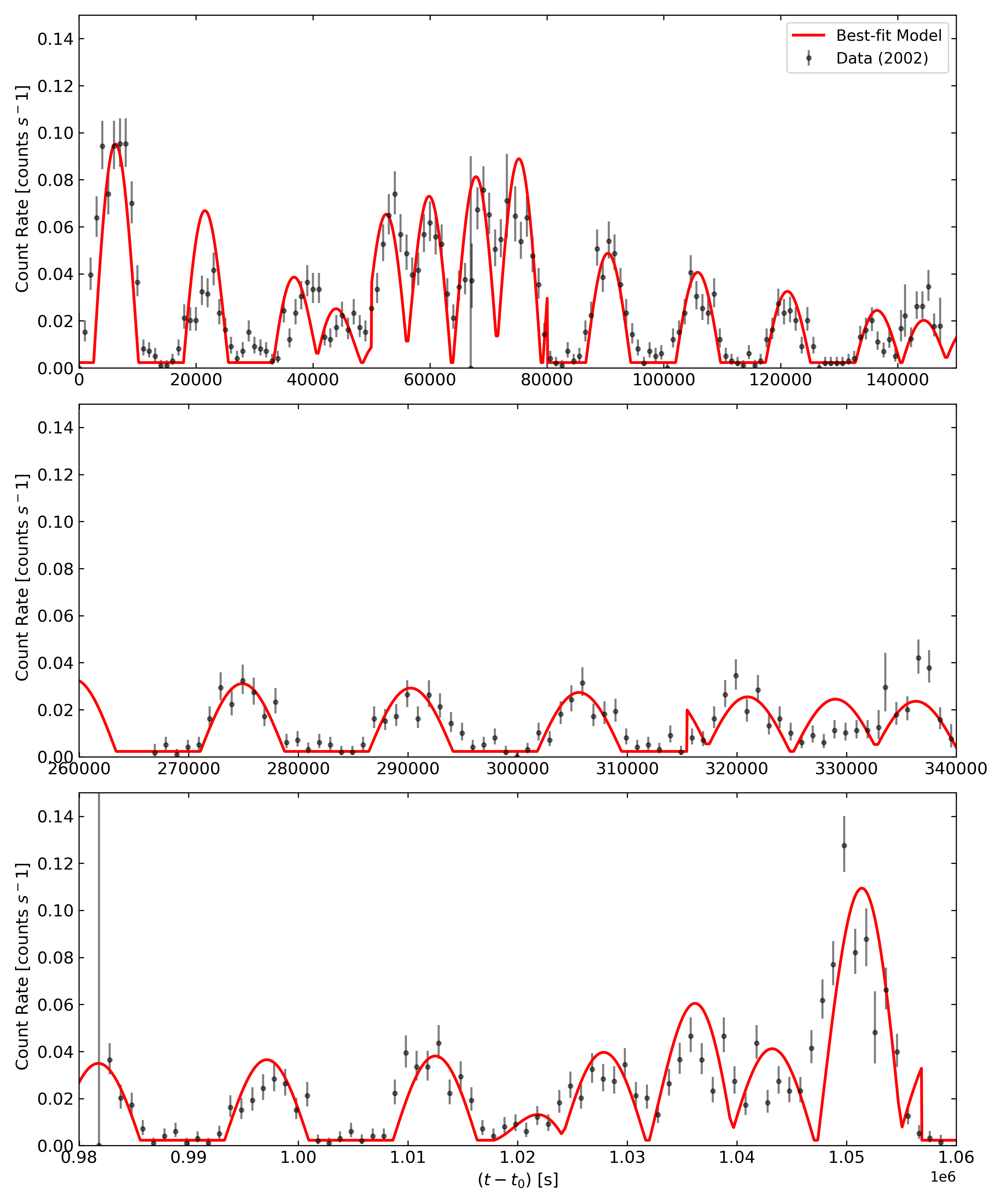}
    \caption{Best-fit model derived from MCMC analysis, divided into the three distinct observing windows from Figure \ref{fig:raw_2002}. Top panel: Merged light curve of observations 2735, 3384, 2736, and 3385. Middle panel: Observations 2737 and 3386. Bottom panel: Observations 2738 and 3387. The red line represents the model constructed from the median posterior parameters. Time is shown relative to the start of the first observation ($t_0$).}
    \label{fig:fitting}
\end{figure}

Figure \ref{fig:fitting} presents the X-ray light curve of the 2002 observation of X10 with the best-fit model constructed using the parameters from Table \ref{tab:full_results}. The model successfully reproduces the complex pulse profile morphology, including the alternating visibility of the accretion poles. 
The visibility functions, $B_1(t)$ and $B_2(t)$, capture the modulation introduced by the beat period, reproducing the observed changes in phase and relative intensity across the different observing epochs. 

In particular, we note the introduction of visibility of the second pole around t=45,000 s, and the disappearance of the first pole around t=80,000 s. Similarly, another pole switch is seen at the end of the last observation, around $t=1.03\times10^6$ s.  A pole switch is also seen around t=145,000 s, successfully reproduced in our model, and a hint of a pole switch beginning around t=325,000 s. 

The total $\chi^2$ of the fit is 1269.4 for 281 degrees of freedom, so the reduced $\chi^2$ is still relatively poor at 4.52. This can be attributed to the stochastic nature of accretion, and perhaps subtleties not accounted for in our model.  Nevertheless, this relatively simple model accurately captures the fundamental features of the lightcurve, suggesting that we have correctly identified the spin and orbital periods of the system.

\subsection{Pulse Timing (O--C)}

\begin{figure}[ht!]
    \centering
    \includegraphics[width=0.7\linewidth]{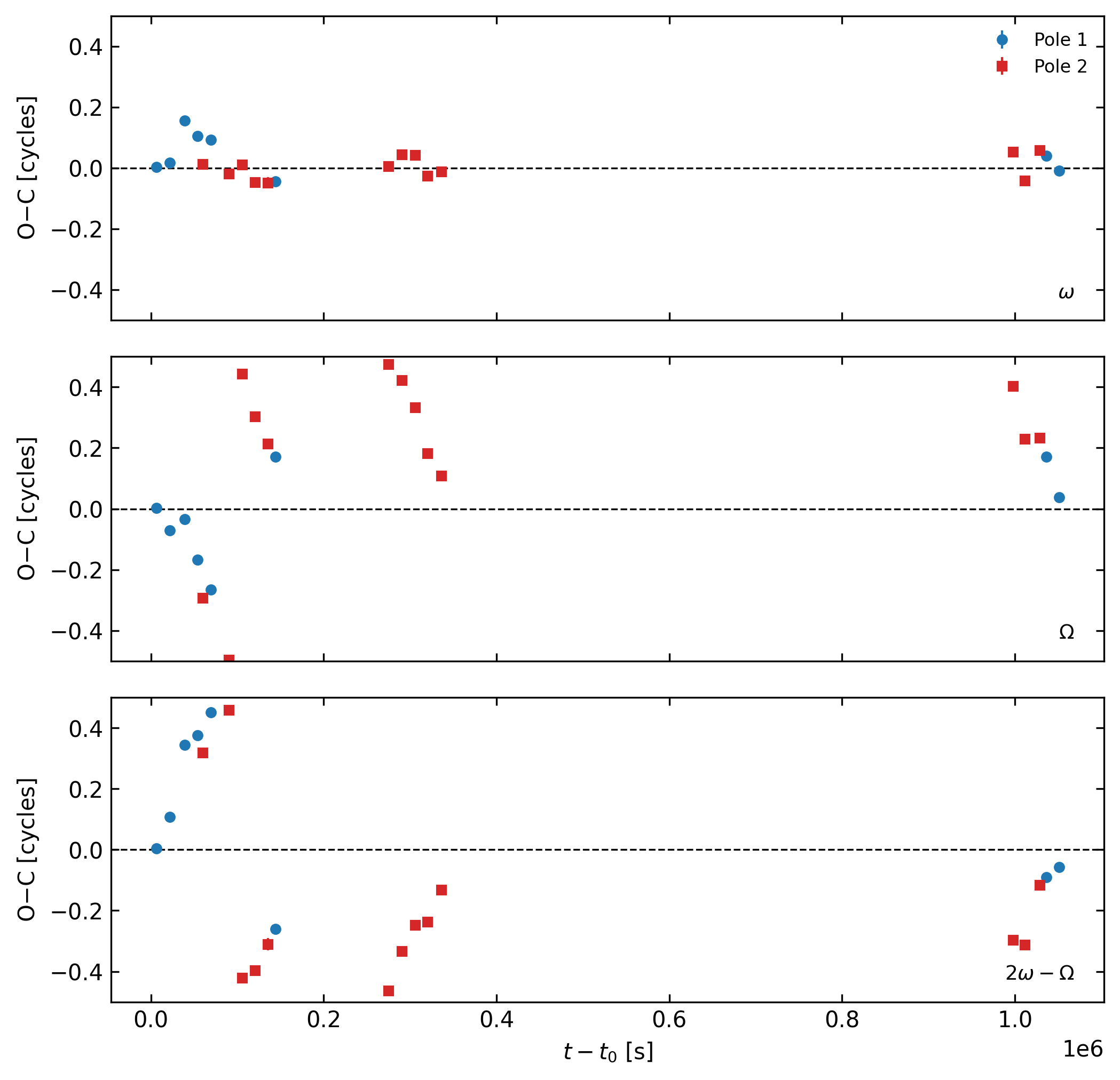}
    \caption{O--C diagram of the X-ray pulse times of X10 for the MCMC spin frequency (top, $\omega$) and the two strongest periodogram peaks, $\Omega$ (middle) and $2\omega-\Omega$ (bottom). Blue circles and red squares mark pulses from pole 1 and pole 2, respectively. The $1\sigma$ uncertainties are smaller than the symbols. Only $\omega$ keeps the pulses near O--C $=0$ in all three observing windows.
    }
    \label{fig:oc}
\end{figure}

To test the period obtained through the MCMC fit independently, we measured the times of all 21 X-ray pulses and compared them with the times predicted by three candidate frequencies. These are the MCMC spin frequency $\omega$ and the two strongest periodogram peaks, $\Omega$ and $2\omega - \Omega$. The resulting O--C diagram is shown in Figure~\ref{fig:oc}. For $\omega$, the pulses from both poles stay close to O--C $=0$ across all three observing windows, with a scatter of only 0.05 cycles. In contrast, for $\Omega$ and $2\omega-\Omega$ the O--C values span the full range from $-0.5$ to $0.5$, with scatters of 0.26 and 0.30 cycles, close to the 0.29 cycles expected for random phases.  These results indicate that the X-ray pulses repeat at $\omega$, and not at either periodogram peak.

\section{Discussion}

\begin{figure}
\centering
\includegraphics[width=\textwidth]{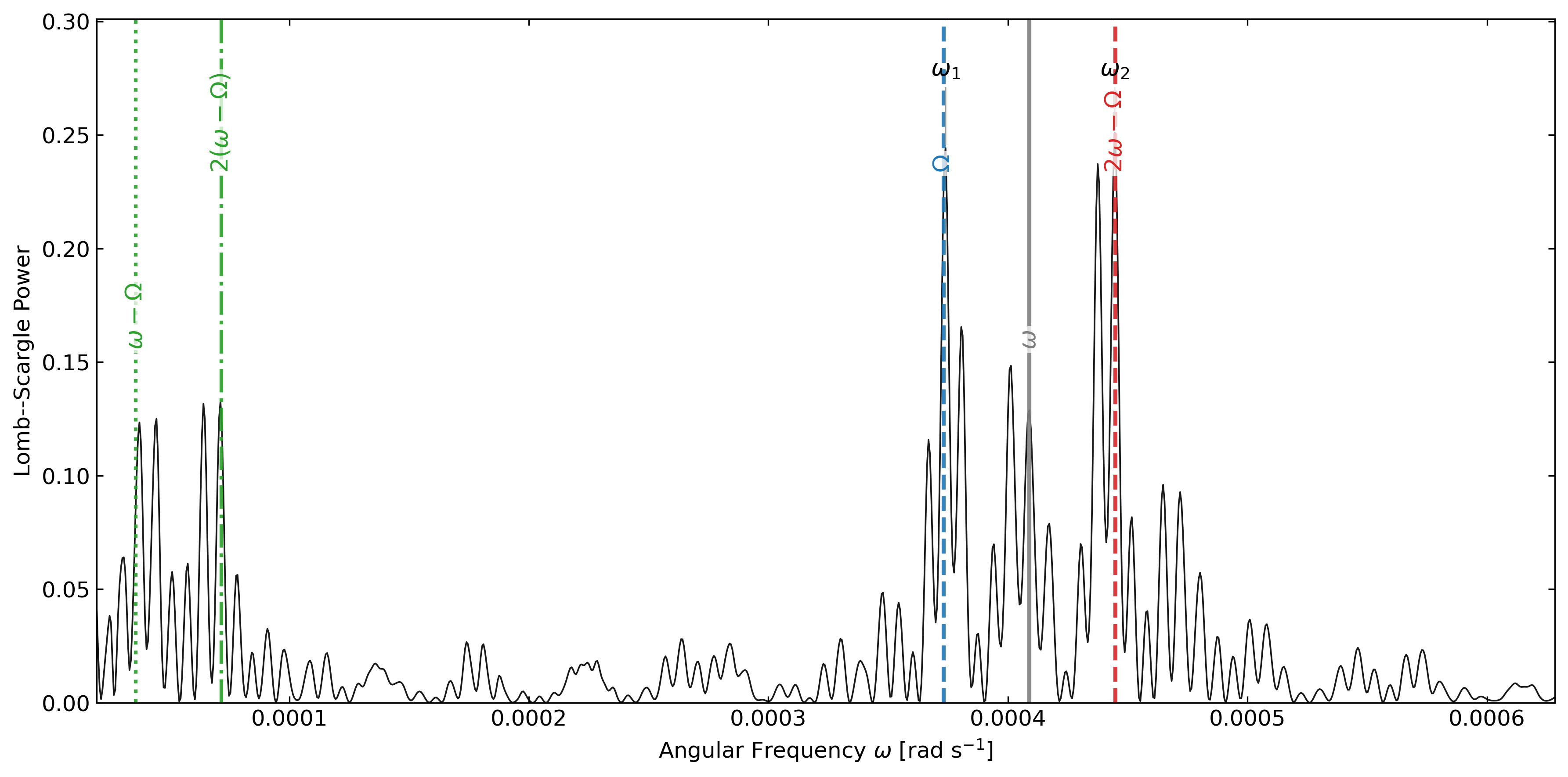}
\caption{Lomb-Scargle periodogram of the 2002 dataset, with the frequencies derived from MCMC model. The two dominant observed peaks, $\omega_1$ and $\omega_2$, coincide with the orbital frequency $\Omega$ (blue dashed) and the sideband $2\omega-\Omega$ (red dashed) derived from the model, while the spin frequency $\omega$ (gray solid) itself lies between them with lower power. The beat frequency $\omega-\Omega$ (green dotted) is likewise suppressed relative to its harmonic $2(\omega-\Omega)$ (green dash-dotted).}
\label{fig:sidebands}
\end{figure}

The results of the MCMC analysis show that the spin angular frequency ($\omega$) of X10 lies almost exactly between the two most prominent peaks identified in the periodogram (Figure \ref{fig:periodogram}), $\omega_1$ and $\omega_2$. This frequency configuration was first addressed by \citet{wynn}, in which the spin signal is suppressed in the power spectrum and is replaced by sideband frequencies when the system goes through a complete pole-switching between the two magnetic poles. In this case, the lower-frequency peak, $\omega_1$, matches the orbital angular frequency ($\Omega$) derived from the numerical modeling, with a difference of only $7.80 \times 10^{-7}$ rad s$^{-1}$ (0.21\%). The second peak, $\omega_2$, according to Wynn \& King's description, corresponds to the sideband frequency  $2\omega - \Omega$. As seen in Figure \ref{fig:sidebands}, the Lomb-Scargle power at $\omega$ (0.13) is roughly half that at $\omega_1$ (0.24) and $\omega_2$ (0.25), so the spin frequency itself produces no distinctive peak despite being the fundamental rotation frequency of the white dwarf.

This scenario was later developed into a predictive "spot model" by  \citet{Wang_2020}. In their terminology, our situation corresponds to a ``Class 2" accretion geometry, equivalent to what 
\citet{halpern_resolving_2024} defines as ``Case B",  where the two magnetic poles alternate completely over the beat cycle, rather than one pole remaining always visible (Class 1/Case A). In Class 2 systems, the beat frequency peak is suppressed along with the spin frequency peak, reappearing instead at its harmonic frequency. In 
our system, 
the value of the beat period obtained through MCMC, $P_{\mathrm{beat}} = 1.7550{\times10^{5}}$ s $(\omega_{\mathrm{beat}} = 3.58 \times10^{-5}$ rad s$^{-1}$), does not correspond with any maximum in the periodogram. Instead, the strongest peak seen is the second harmonic, $2\omega_{\mathrm{beat}} = 7.16 \times 10^{-5}$ rad s$^{-1}$, closely matching the predictions of \citet{Wang_2020}. They found a similar absence of the beat period in their 
application of the spot model to  a $\sim1.47$ hr peak in 1RXS J083842.1-282723's 
XMM-Newton periodogram, originally attributed by \citet{Rea17} to the spin period, indicating instead that it was a higher-order sideband of the spin and orbital frequencies rather than the spin period itself. 

A similar case among asynchronous polars was reported by \citet{littlefield_fast-cadence_2019}, who showed that the long-accepted spin 
period of 
CD Ind had in fact been confused with a sideband, $2\omega-\Omega$. Pole-switching causes the algorithm to find whichever frequency keeps the light curve most stable in phase, which is not necessarily the true rotation frequency of the white dwarf. In our case, the same effect suppresses both $\omega$ and $\omega_{\mathrm{beat}}$ from the periodogram of X10.

Our results 
verify the orbital period previously suggested for X10, while showing that the spin period is different. 
The derived value of $P_{\mathrm{orb}} = 4.68$ hours is 
consistent with the $\sim4.67$ hour modulation 
suggested by 
\citet{Heinke2005}, and independently measured by \citet{Bao23}. This agreement 
indicates 
that the dominant signal observed in the 2002 dataset corresponds to the orbital motion rather than the white dwarf spin. 
The derived spin period, $P_{\mathrm{spin}} = 4.27$ hours, 
supports the classification of X10 as an asynchronous polar.

We note that our derived orbital period of 4.68 hours is one of the longest known for a polar CV (along with 4.64 hours for SDSS J084617.11+245344.1, \citealt{Littlefield23b},  8 hours for V1309 Ori, \citealt{Garnavich94}, and V479 Andromedae at 14.25 hours, \citealt{GonzalezBuitrago13}; see also the candidate polars 47 Tuc W20 and W23, \citealt{Bao23}). This long orbital period naturally predicts a high mass transfer rate, and thus its relatively high X-ray luminosity ($L_X\sim2.2\times10^{32}$ erg s$^{-1}$)compared to most other polar CVs (typically $10^{30}--10^{32}$ erg s$^{-1}$; see \citealt{Schwope24}). However, X10's $L_X$ is well below the expectation (from, e.g., \citealt{Howell01}) for its accretion luminosity at this orbital period, $L_X=GM\dot{M}/R_{wd} \sim 3\times10^{34}$ erg s$^{-1}$, assuming a 0.6 $M_{\odot}$ WD. While some of the missing accretion luminosity may be emerging in the far-UV as a softer blackbody component \citep{cropper90}, it has also been noted that long-period polars often have mass accretion rates much lower than other CVs at the same orbital periods, suggesting that magnetic braking is reduced, perhaps due to the strong B field of the primary \citep{AraujoBetancor05}.

From our results, we estimate a degree of synchronism of $P_{\mathrm{spin}}/P_{\mathrm{orb}} =  0.9124 \pm 0.0002$. 
The $P_{\mathrm{spin}}$ is thus significantly farther from $P_{\mathrm{orb}}$ 
than the typical $<2\%$ observed in near-synchronous polars such as V1500~Cyg \citep{Schwarz07}. X10 would belong to the uncommon subgroup of asynchronous CVs with $P_{\mathrm{spin}}/P_{\mathrm{orb}}$ ratios between 0.87 and 0.97, such as Paloma \citep{Schwarz07}, IGR J19552+0044 \citep{Tovmassian17}, and V381 Vel \citep{norton_spin_2004}. Theoretical models of rotational equilibria by \citet{norton_spin_2004} indicate that intermediate polars with $P_{\mathrm{spin}}/P_{\mathrm{orb}}>0.6$ cannot maintain stable asynchronous rotation and are headed to synchronism, suggesting that X10 may represent a transitional system between an intermediate polar and a fully synchronized polar.

\section{Conclusion}

In this work, we performed a detailed timing analysis of the magnetic cataclysmic variable X10 in the globular cluster 47 Tucanae using Chandra X-ray observations. By focusing on the 2002 dataset, which exhibits the clearest pulsations and pole-switching behavior, we applied a physically motivated sinusoidal model combined with Markov Chain Monte Carlo (MCMC) sampling to find the intrinsic spin, orbital, and beat periodicities of the system.

By modeling the X-ray light curve with two rotating accretion poles whose visibility is modulated by the beat period, and by adopting a Poisson likelihood appropriate for low-count X-ray data, we constrained the spin period of the white dwarf to $P_{\mathrm{spin}} = 4.2683^{+0.0001}_{-0.0002} \, \mathrm{ h} $. The beat modulation was found to have a period of
$P_{\mathrm{beat}} = 48.75^{+0.07}_{-0.10} \, \mathrm{ h}$, which allowed us to derive an orbital period of 
$P_{\mathrm{orb}} = 4.6779^{+0.0010}_{-0.0007} \,\mathrm{h}$. 

These results demonstrate that the dominant periodic signal observed in the 2002 X-ray data corresponds to the orbital motion rather than the white dwarf spin. The true spin frequency lies between the two strongest peaks in the Lomb–Scargle periodogram and is suppressed by sideband modulation, a behaviour consistent with the “Case B” accretion scenario described by \cite{halpern_resolving_2024}. 

Our derived orbital period is consistent with the modulation originally reported by \cite{Heinke2005}, 
supporting the classification of X10 as an asynchronous polar of unusually long period. The inferred degree of asynchronism, 8.8\%, is higher than typically observed in near-synchronous polars, suggesting that X10 may represent a transitional system evolving toward synchronism.

Future analysis could incorporate the lower S/N  light curves from the 2000, 2005, 2014, and 2021 Chandra observations to further constrain X10’s accretion geometry and evolutionary state.

\begin{acknowledgments}
C. O. Heinke is supported by NSERC Discovery Grant RGPIN-2023-04264. We thank Peter Edmonds and Josh Grindlay for discussions over many years. 

\end{acknowledgments}




%
\facilities{CXO (ACIS)}

\software{astropy \citep{Astropy13,Astropy18,Astropy22}, emcee \citep{foreman-mackey_emcee_2013}, SciPy \citep{2020SciPy-NMeth} , CIAO \citep{Fruscione26} 
    }      


\appendix


\begin{figure}[ht!]
    \centering
    \includegraphics[width=1.0\linewidth]{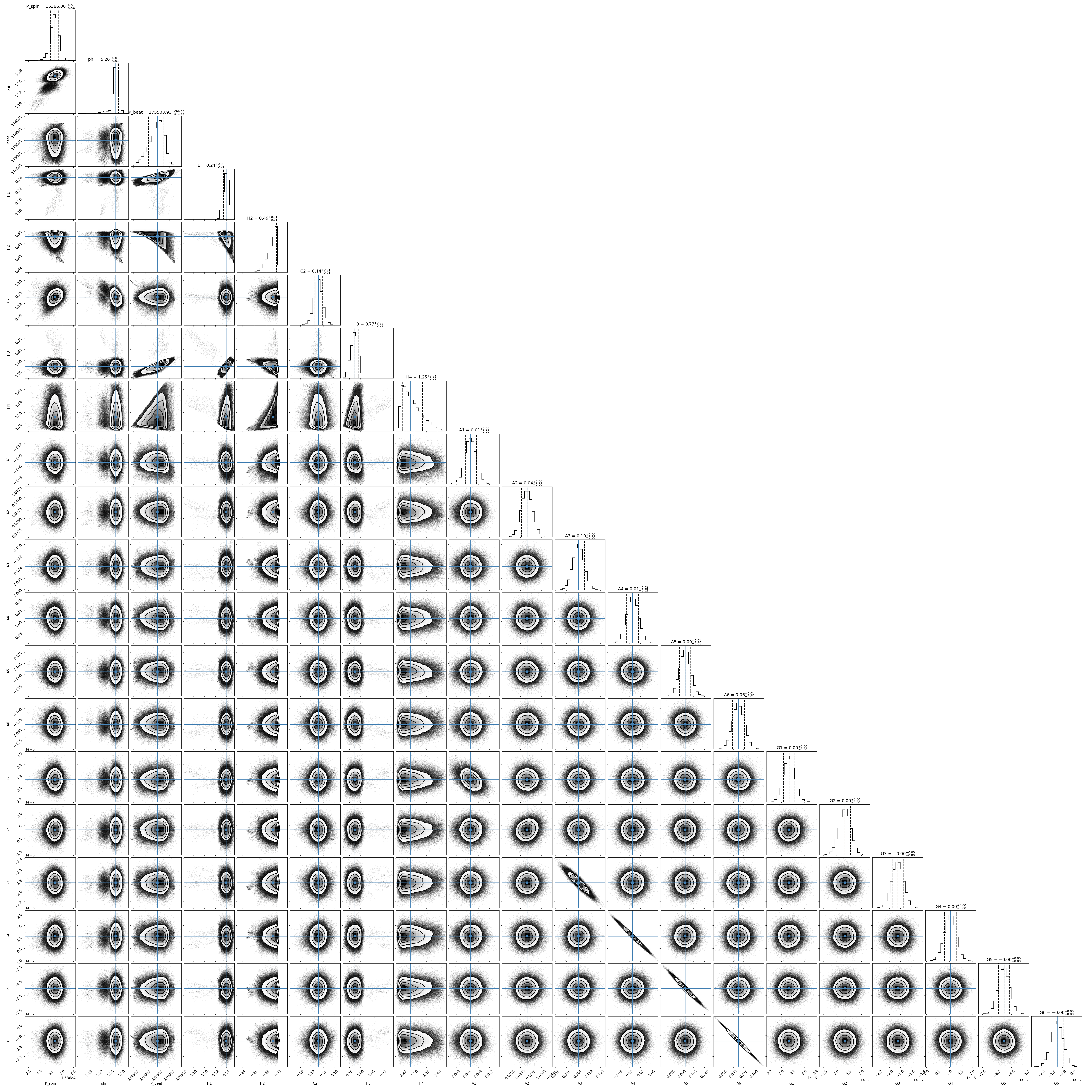}
    \caption{Corner plot showing the marginalized posterior distributions of the model parameters obtained from the MCMC analysis. The one-dimensional histograms represent the marginalized posterior for each parameter, while the two-dimensional contours show the 68\% and 95\% credible regions. Vertical dashed lines indicate the median values of the posterior distributions. 
    }
    \label{fig:corner}
\end{figure}


\bibliography{CV}{}

\begin{thebibliography}{}
\expandafter\ifx\csname natexlab\endcsname\relax\def\natexlab#1{#1}\fi
\providecommand{\url}[1]{\href{#1}{#1}}
\providecommand{\dodoi}[1]{doi:~\href{http://doi.org/#1}{\nolinkurl{#1}}}
\providecommand{\doeprint}[1]{\href{http://ascl.net/#1}{\nolinkurl{http://ascl.net/#1}}}
\providecommand{\doarXiv}[1]{\href{https://arxiv.org/abs/#1}{\nolinkurl{https://arxiv.org/abs/#1}}}

\bibitem[{R. {Amato} {et~al.}(2024){Amato}, {La Palombara}, {Imbrogno}, {Israel}, {Esposito}, {de Martino}, {Webb}, \& {Iaria}}]{Amato24}
{Amato}, R., {La Palombara}, N., {Imbrogno}, M., {et~al.} 2024, \bibinfo{title}{{Investigating the nature of the 2.4 h-period eclipsing cataclysmic variable W2 in 47 Tuc},} \aap, 689, A176, \dodoi{10.1051/0004-6361/202450034}

\bibitem[{S. {Araujo-Betancor} {et~al.}(2005){Araujo-Betancor}, {G{\"a}nsicke}, {Long}, {Beuermann}, {de Martino}, {Sion}, \& {Szkody}}]{AraujoBetancor05}
{Araujo-Betancor}, S., {G{\"a}nsicke}, B.~T., {Long}, K.~S., {et~al.} 2005, \bibinfo{title}{{Far-Ultraviolet Spectroscopy of Magnetic Cataclysmic Variables},} \apj, 622, 589, \dodoi{10.1086/427914}

\bibitem[{ {Astropy Collaboration} {et~al.}(2013){Astropy Collaboration}, {Robitaille}, {Tollerud}, {Greenfield}, {Droettboom}, {Bray}, {Aldcroft}, {Davis}, {Ginsburg}, {Price-Whelan}, {Kerzendorf}, {Conley}, {Crighton}, {Barbary}, {Muna}, {Ferguson}, {Grollier}, {Parikh}, {Nair}, {Unther}, {Deil}, {Woillez}, {Conseil}, {Kramer}, {Turner}, {Singer}, {Fox}, {Weaver}, {Zabalza}, {Edwards}, {Azalee Bostroem}, {Burke}, {Casey}, {Crawford}, {Dencheva}, {Ely}, {Jenness}, {Labrie}, {Lim}, {Pierfederici}, {Pontzen}, {Ptak}, {Refsdal}, {Servillat}, \& {Streicher}}]{Astropy13}
{Astropy Collaboration}, {Robitaille}, T.~P., {Tollerud}, E.~J., {et~al.} 2013, \bibinfo{title}{{Astropy: A community Python package for astronomy},} \aap, 558, A33, \dodoi{10.1051/0004-6361/201322068}

\bibitem[{ {Astropy Collaboration} {et~al.}(2018){Astropy Collaboration}, {Price-Whelan}, {Sip{\H{o}}cz}, {G{\"u}nther}, {Lim}, {Crawford}, {Conseil}, {Shupe}, {Craig}, {Dencheva}, {Ginsburg}, {VanderPlas}, {Bradley}, {P{\'e}rez-Su{\'a}rez}, {de Val-Borro}, {Aldcroft}, {Cruz}, {Robitaille}, {Tollerud}, {Ardelean}, {Babej}, {Bach}, {Bachetti}, {Bakanov}, {Bamford}, {Barentsen}, {Barmby}, {Baumbach}, {Berry}, {Biscani}, {Boquien}, {Bostroem}, {Bouma}, {Brammer}, {Bray}, {Breytenbach}, {Buddelmeijer}, {Burke}, {Calderone}, {Cano Rodr{\'\i}guez}, {Cara}, {Cardoso}, {Cheedella}, {Copin}, {Corrales}, {Crichton}, {D'Avella}, {Deil}, {Depagne}, {Dietrich}, {Donath}, {Droettboom}, {Earl}, {Erben}, {Fabbro}, {Ferreira}, {Finethy}, {Fox}, {Garrison}, {Gibbons}, {Goldstein}, {Gommers}, {Greco}, {Greenfield}, {Groener}, {Grollier}, {Hagen}, {Hirst}, {Homeier}, {Horton}, {Hosseinzadeh}, {Hu}, {Hunkeler}, {Ivezi{\'c}}, {Jain}, {Jenness}, {Kanarek}, {Kendrew}, {Kern}, {Kerzendorf}, {Khvalko}, {King}, {Kirkby}, {Kulkarni},
  {Kumar}, {Lee}, {Lenz}, {Littlefair}, {Ma}, {Macleod}, {Mastropietro}, {McCully}, {Montagnac}, {Morris}, {Mueller}, {Mumford}, {Muna}, {Murphy}, {Nelson}, {Nguyen}, {Ninan}, {N{\"o}the}, {Ogaz}, {Oh}, {Parejko}, {Parley}, {Pascual}, {Patil}, {Patil}, {Plunkett}, {Prochaska}, {Rastogi}, {Reddy Janga}, {Sabater}, {Sakurikar}, {Seifert}, {Sherbert}, {Sherwood-Taylor}, {Shih}, {Sick}, {Silbiger}, {Singanamalla}, {Singer}, {Sladen}, {Sooley}, {Sornarajah}, {Streicher}, {Teuben}, {Thomas}, {Tremblay}, {Turner}, {Terr{\'o}n}, {van Kerkwijk}, {de la Vega}, {Watkins}, {Weaver}, {Whitmore}, {Woillez}, {Zabalza}, \& {Astropy Contributors}}]{Astropy18}
{Astropy Collaboration}, {Price-Whelan}, A.~M., {Sip{\H{o}}cz}, B.~M., {et~al.} 2018, \bibinfo{title}{{The Astropy Project: Building an Open-science Project and Status of the v2.0 Core Package},} \aj, 156, 123, \dodoi{10.3847/1538-3881/aabc4f}

\bibitem[{ {Astropy Collaboration} {et~al.}(2022){Astropy Collaboration}, {Price-Whelan}, {Lim}, {Earl}, {Starkman}, {Bradley}, {Shupe}, {Patil}, {Corrales}, {Brasseur}, {N{\"o}the}, {Donath}, {Tollerud}, {Morris}, {Ginsburg}, {Vaher}, {Weaver}, {Tocknell}, {Jamieson}, {van Kerkwijk}, {Robitaille}, {Merry}, {Bachetti}, {G{\"u}nther}, {Aldcroft}, {Alvarado-Montes}, {Archibald}, {B{\'o}di}, {Bapat}, {Barentsen}, {Baz{\'a}n}, {Biswas}, {Boquien}, {Burke}, {Cara}, {Cara}, {Conroy}, {Conseil}, {Craig}, {Cross}, {Cruz}, {D'Eugenio}, {Dencheva}, {Devillepoix}, {Dietrich}, {Eigenbrot}, {Erben}, {Ferreira}, {Foreman-Mackey}, {Fox}, {Freij}, {Garg}, {Geda}, {Glattly}, {Gondhalekar}, {Gordon}, {Grant}, {Greenfield}, {Groener}, {Guest}, {Gurovich}, {Handberg}, {Hart}, {Hatfield-Dodds}, {Homeier}, {Hosseinzadeh}, {Jenness}, {Jones}, {Joseph}, {Kalmbach}, {Karamehmetoglu}, {Ka{\l}uszy{\'n}ski}, {Kelley}, {Kern}, {Kerzendorf}, {Koch}, {Kulumani}, {Lee}, {Ly}, {Ma}, {MacBride}, {Maljaars}, {Muna}, {Murphy}, {Norman},
  {O'Steen}, {Oman}, {Pacifici}, {Pascual}, {Pascual-Granado}, {Patil}, {Perren}, {Pickering}, {Rastogi}, {Roulston}, {Ryan}, {Rykoff}, {Sabater}, {Sakurikar}, {Salgado}, {Sanghi}, {Saunders}, {Savchenko}, {Schwardt}, {Seifert-Eckert}, {Shih}, {Jain}, {Shukla}, {Sick}, {Simpson}, {Singanamalla}, {Singer}, {Singhal}, {Sinha}, {Sip{\H{o}}cz}, {Spitler}, {Stansby}, {Streicher}, {{\v{S}}umak}, {Swinbank}, {Taranu}, {Tewary}, {Tremblay}, {de Val-Borro}, {Van Kooten}, {Vasovi{\'c}}, {Verma}, {de Miranda Cardoso}, {Williams}, {Wilson}, {Winkel}, {Wood-Vasey}, {Xue}, {Yoachim}, {Zhang}, {Zonca}, \& {Astropy Project Contributors}}]{Astropy22}
{Astropy Collaboration}, {Price-Whelan}, A.~M., {Lim}, P.~L., {et~al.} 2022, \bibinfo{title}{{The Astropy Project: Sustaining and Growing a Community-oriented Open-source Project and the Latest Major Release (v5.0) of the Core Package},} \apj, 935, 167, \dodoi{10.3847/1538-4357/ac7c74}

\bibitem[{A. {Bahramian} {et~al.}(2020){Bahramian}, {Strader}, {Miller-Jones}, {Chomiuk}, {Heinke}, {Maccarone}, {Pooley}, {Shishkovsky}, {Tudor}, {Zhao}, {Li}, {Sivakoff}, {Tremou}, \& {Buchner}}]{Bahramian2020}
{Bahramian}, A., {Strader}, J., {Miller-Jones}, J. C.~A., {et~al.} 2020, \bibinfo{title}{{The MAVERIC Survey: Chandra/ACIS Catalog of Faint X-Ray Sources in 38 Galactic Globular Clusters},} \apj, 901, 57, \dodoi{10.3847/1538-4357/aba51d}

\bibitem[{T. {Bao} {et~al.}(2023){Bao}, {Li}, \& {Cheng}}]{Bao23}
{Bao}, T., {Li}, Z., \& {Cheng}, Z. 2023, \bibinfo{title}{{Periodic X-ray sources in the massive globular cluster 47 Tucanae: Evidence for dynamically formed cataclysmic variables},} \mnras, 521, 4257, \dodoi{10.1093/mnras/stad836}

\bibitem[{T. {Bao} {et~al.}(2024){Bao}, {Li}, {Cheng}, \& {Belloni}}]{Bao24}
{Bao}, T., {Li}, Z., {Cheng}, Z., \& {Belloni}, D. 2024, \bibinfo{title}{{A Chandra survey of Milky Way globular clusters - IV. Periodic X-ray sources},} \mnras, 527, 7173, \dodoi{10.1093/mnras/stad3665}

\bibitem[{D. {Belloni} {et~al.}(2021){Belloni}, {Schreiber}, {Salaris}, {Maccarone}, \& {Zorotovic}}]{Belloni21}
{Belloni}, D., {Schreiber}, M.~R., {Salaris}, M., {Maccarone}, T.~J., \& {Zorotovic}, M. 2021, \bibinfo{title}{{Magnetic dynamos in white dwarfs - I. Explaining the dearth of bright intermediate polars in globular clusters},} \mnras, 505, L74, \dodoi{10.1093/mnrasl/slab054}

\bibitem[{F. {Bernardini} {et~al.}(2013){Bernardini}, {de Martino}, {Mukai}, {Falanga}, {Andruchow}, {Bonnet-Bidaud}, {Masetti}, {Buitrago}, {Mouchet}, \& {Tovmassian}}]{Bernardini13}
{Bernardini}, F., {de Martino}, D., {Mukai}, K., {et~al.} 2013, \bibinfo{title}{{On the nature of the hard X-ray sources SWIFT J1907.3-2050, IGR J12123-5802 and IGR J19552+0044},} \mnras, 435, 2822, \dodoi{10.1093/mnras/stt1434}

\bibitem[{C.~G. {Campbell} \& A.~D. {Schwope}(1999){Campbell} \& {Schwope}}]{Campbell99}
{Campbell}, C.~G., \& {Schwope}, A.~D. 1999, \bibinfo{title}{{Asynchronous rotation in the polars},} \aap, 343, 132

\bibitem[{W. {Cash}(1979){Cash}}]{Cash79}
{Cash}, W. 1979, \bibinfo{title}{{Parameter estimation in astronomy through application of the likelihood ratio.},} \apj, 228, 939, \dodoi{10.1086/156922}

\bibitem[{S. {Chen} {et~al.}(2018){Chen}, {Richer}, {Caiazzo}, \& {Heyl}}]{Chen18}
{Chen}, S., {Richer}, H., {Caiazzo}, I., \& {Heyl}, J. 2018, \bibinfo{title}{{Distances to the Globular Clusters 47 Tucanae and NGC 362 Using Gaia DR2 Parallaxes},} \apj, 867, 132, \dodoi{10.3847/1538-4357/aae089}

\bibitem[{M. {Cropper}(1990){Cropper}}]{cropper90}
{Cropper}, M. 1990, \bibinfo{title}{{The Polars},} \ssr, 54, 195, \dodoi{10.1007/BF00177799}

\bibitem[{A. {Dobrotka} {et~al.}(2006){Dobrotka}, {Lasota}, \& {Menou}}]{Dobrotka06}
{Dobrotka}, A., {Lasota}, J.-P., \& {Menou}, K. 2006, \bibinfo{title}{{All Quiet in Globular Clusters},} \apj, 640, 288, \dodoi{10.1086/500042}

\bibitem[{P.~D. {Edmonds} {et~al.}(2003{\natexlab{a}}){Edmonds}, {Gilliland}, {Heinke}, \& {Grindlay}}]{Edmonds03a}
{Edmonds}, P.~D., {Gilliland}, R.~L., {Heinke}, C.~O., \& {Grindlay}, J.~E. 2003{\natexlab{a}}, \bibinfo{title}{{An Extensive Census of Hubble Space Telescope Counterparts to Chandra X-Ray Sources in the Globular Cluster 47 Tucanae. I. Astrometry and Photometry},} \apj, 596, 1177, \dodoi{10.1086/378193}

\bibitem[{P.~D. {Edmonds} {et~al.}(2003{\natexlab{b}}){Edmonds}, {Gilliland}, {Heinke}, \& {Grindlay}}]{Edmonds03b}
{Edmonds}, P.~D., {Gilliland}, R.~L., {Heinke}, C.~O., \& {Grindlay}, J.~E. 2003{\natexlab{b}}, \bibinfo{title}{{An Extensive Census of Hubble Space Telescope Counterparts to Chandra X-Ray Sources in the Globular Cluster 47 Tucanae. II. Time Series and Analysis},} \apj, 596, 1197, \dodoi{10.1086/378194}

\bibitem[{P.~D. {Edmonds} {et~al.}(1999){Edmonds}, {Grindlay}, {Cool}, {Cohn}, {Lugger}, \& {Bailyn}}]{Edmonds99}
{Edmonds}, P.~D., {Grindlay}, J.~E., {Cool}, A., {et~al.} 1999, \bibinfo{title}{{Cataclysmic Variables and a Candidate Helium White Dwarf in the Globular Cluster NGC 6397},} \apj, 516, 250, \dodoi{10.1086/307106}

\bibitem[{D. Foreman-Mackey {et~al.}(2013{\natexlab{a}})Foreman-Mackey, Hogg, Lang, \& Goodman}]{Foreman_Mackey_2013}
Foreman-Mackey, D., Hogg, D.~W., Lang, D., \& Goodman, J. 2013{\natexlab{a}}, \bibinfo{title}{<tt>emcee</tt>: The MCMC Hammer,} Publications of the Astronomical Society of the Pacific, 125, 306–312, \dodoi{10.1086/670067}

\bibitem[{D. Foreman-Mackey {et~al.}(2013{\natexlab{b}})Foreman-Mackey, Hogg, Lang, \& Goodman}]{foreman-mackey_emcee_2013}
Foreman-Mackey, D., Hogg, D.~W., Lang, D., \& Goodman, J. 2013{\natexlab{b}}, \bibinfo{title}{emcee: {The} {MCMC} {Hammer},} Publications of the Astronomical Society of the Pacific, 125, 306, \dodoi{10.1086/670067}

\bibitem[{A. {Fruscione} {et~al.}(2026){Fruscione}, {McDowell}, {Burke}, {Cresitello-Dittmar}, {Evans}, {Evans}, {Glotfelty}, {G{\"u}nther}, {Huenemoerder}, {Joye}, {Lee}, {McLaughlin}, {Miller}, {Nynka}, {Principe}, \& {Siemiginowska}}]{Fruscione26}
{Fruscione}, A., {McDowell}, J.~C., {Burke}, D.~J., {et~al.} 2026, \bibinfo{title}{{CIAO: Chandra's Data Analysis System for X-Ray Astronomy and Beyond},} \apj, 1005, 116, \dodoi{10.3847/1538-4357/ae6db4}

\bibitem[{G.~P. {Garmire} {et~al.}(2003){Garmire}, {Bautz}, {Ford}, {Nousek}, \& {Ricker}}]{garmire2003}
{Garmire}, G.~P., {Bautz}, M.~W., {Ford}, P.~G., {Nousek}, J.~A., \& {Ricker}, Jr., G.~R. 2003, \bibinfo{title}{{Advanced CCD imaging spectrometer (ACIS) instrument on the Chandra X-ray Observatory},} in Society of Photo-Optical Instrumentation Engineers (SPIE) Conference Series, Vol. 4851, X-Ray and Gamma-Ray Telescopes and Instruments for Astronomy., ed. J.~E. {Truemper} \& H.~D. {Tananbaum}, 28--44, \dodoi{10.1117/12.461599}

\bibitem[{P.~M. {Garnavich} {et~al.}(1994){Garnavich}, {Szkody}, {Robb}, {Zurek}, \& {Hoard}}]{Garnavich94}
{Garnavich}, P.~M., {Szkody}, P., {Robb}, R.~M., {Zurek}, D.~R., \& {Hoard}, D.~W. 1994, \bibinfo{title}{{RJ 051542+0104.7: A New Magnetic, Eclipsing, Cataclysmic Variable},} \apjl, 435, L141, \dodoi{10.1086/187614}

\bibitem[{N. {Gehrels}(1986){Gehrels}}]{Gehrels86}
{Gehrels}, N. 1986, \bibinfo{title}{{Confidence Limits for Small Numbers of Events in Astrophysical Data},} \apj, 303, 336, \dodoi{10.1086/164079}

\bibitem[{D. {Gonz{\'a}lez-Buitrago} {et~al.}(2013){Gonz{\'a}lez-Buitrago}, {Tovmassian}, {Zharikov}, {Yungelson}, {Miyaji}, {Echevarr{\'\i}a}, {Aviles}, \& {Valyavin}}]{GonzalezBuitrago13}
{Gonz{\'a}lez-Buitrago}, D., {Tovmassian}, G., {Zharikov}, S., {et~al.} 2013, \bibinfo{title}{{Multiwavelength observations of V479 Andromedae: a close compact binary with an identity crisis},} \aap, 553, A28, \dodoi{10.1051/0004-6361/201219939}

\bibitem[{J.~E. {Grindlay} {et~al.}(1995){Grindlay}, {Cool}, {Callanan}, {Bailyn}, {Cohn}, \& {Lugger}}]{Grindlay95}
{Grindlay}, J.~E., {Cool}, A.~M., {Callanan}, P.~J., {et~al.} 1995, \bibinfo{title}{{Spectroscopic Identification of Probable Cataclysmic Variables in the Globular Cluster NGC 6397},} \apjl, 455, L47, \dodoi{10.1086/309806}

\bibitem[{J.~E. Grindlay {et~al.}(2001)Grindlay, Heinke, Edmonds, \& Murray}]{Grindlay_2001}
Grindlay, J.~E., Heinke, C., Edmonds, P.~D., \& Murray, S.~S. 2001, \bibinfo{title}{High-Resolution X-ray Imaging of a Globular Cluster Core: Compact Binaries in 47Tuc,} Science, 292, 2290–2295, \dodoi{10.1126/science.1061135}

\bibitem[{J.~P. {Halpern}(2022){Halpern}}]{Halpern22}
{Halpern}, J.~P. 2022, \bibinfo{title}{{Swift J0503.7-2819: a Short-period Asynchronous Polar or Stream-fed Intermediate Polar},} \apj, 934, 123, \dodoi{10.3847/1538-4357/ac7d50}

\bibitem[{J.~P. Halpern(2024)Halpern}]{halpern_resolving_2024}
Halpern, J.~P. 2024, \bibinfo{title}{Resolving the {Periods} of the {Asynchronous} {Polar} {1RXS} {J083842}.1─282723,} The Astrophysical Journal, 963, 78, \dodoi{10.3847/1538-4357/ad24ed}

\bibitem[{J.~P. {Halpern} {et~al.}(2017){Halpern}, {Bogdanov}, \& {Thorstensen}}]{Halpern17}
{Halpern}, J.~P., {Bogdanov}, S., \& {Thorstensen}, J.~R. 2017, \bibinfo{title}{{X-Ray and Optical Study of the Gamma-ray Source 3FGL J0838.8-2829: Identification of a Candidate Millisecond Pulsar Binary and an Asynchronous Polar},} \apj, 838, 124, \dodoi{10.3847/1538-4357/838/2/124}

\bibitem[{W.~E. {Harris}(2010){Harris}}]{Harris10}
{Harris}, W.~E. 2010, \bibinfo{title}{{A New Catalog of Globular Clusters in the Milky Way},} arXiv e-prints, arXiv:1012.3224, \dodoi{10.48550/arXiv.1012.3224}

\bibitem[{G. {Hasinger} {et~al.}(1994){Hasinger}, {Johnston}, \& {Verbunt}}]{Hasinger94}
{Hasinger}, G., {Johnston}, H.~M., \& {Verbunt}, F. 1994, \bibinfo{title}{{Discovery of multiple X-ray sources in 47 Tucanae},} \aap, 288, 466

\bibitem[{A. Hattawi(2019)Hattawi}]{hattawi2019x10}
Hattawi, A. 2019, Master's thesis, University of Alberta, \dodoi{10.7939/r3-tt6h-yp36}

\bibitem[{C.~O. {Heinke} {et~al.}(2005){Heinke}, {Grindlay}, {Edmonds}, {Cohn}, {Lugger}, {Camilo}, {Bogdanov}, \& {Freire}}]{Heinke2005}
{Heinke}, C.~O., {Grindlay}, J.~E., {Edmonds}, P.~D., {et~al.} 2005, \bibinfo{title}{{A Deep Chandra Survey of the Globular Cluster 47 Tucanae: Catalog of Point Sources},} \apj, 625, 796, \dodoi{10.1086/429899}

\bibitem[{C.~O. {Heinke} {et~al.}(2008){Heinke}, {Ruiter}, {Muno}, \& {Belczynski}}]{Heinke08}
{Heinke}, C.~O., {Ruiter}, A.~J., {Muno}, M.~P., \& {Belczynski}, K. 2008, \bibinfo{title}{{Cataclysmic Variables in Globular Clusters, the Galactic Center, and Local Space},} in American Institute of Physics Conference Series, Vol. 1010, A Population Explosion: The Nature \& Evolution of X-ray Binaries in Diverse Environments, ed. R.~M. {Bandyopadhyay}, S.~{Wachter}, D.~{Gelino}, \& C.~R. {Gelino} (AIP), 136--142, \dodoi{10.1063/1.2945021}

\bibitem[{S.~B. {Howell} {et~al.}(2001){Howell}, {Nelson}, \& {Rappaport}}]{Howell01}
{Howell}, S.~B., {Nelson}, L.~A., \& {Rappaport}, S. 2001, \bibinfo{title}{{An Exploration of the Paradigm for the 2-3 Hour Period Gap in Cataclysmic Variables},} \apj, 550, 897, \dodoi{10.1086/319776}

\bibitem[{N. {Ivanova} {et~al.}(2006){Ivanova}, {Heinke}, {Rasio}, {Taam}, {Belczynski}, \& {Fregeau}}]{Ivanova06}
{Ivanova}, N., {Heinke}, C.~O., {Rasio}, F.~A., {et~al.} 2006, \bibinfo{title}{{Formation and evolution of compact binaries in globular clusters - I. Binaries with white dwarfs},} \mnras, 372, 1043, \dodoi{10.1111/j.1365-2966.2006.10876.x}

\bibitem[{A. {Joshi} {et~al.}(2016){Joshi}, {Pandey}, {Singh}, \& {Agrawal}}]{Joshi16}
{Joshi}, A., {Pandey}, J.~C., {Singh}, K.~P., \& {Agrawal}, P.~C. 2016, \bibinfo{title}{{PALOMA: A Magnetic CV between Polars and Intermediate Polars},} \apj, 830, 56, \dodoi{10.3847/0004-637X/830/2/56}

\bibitem[{C. {Knigge}(2012){Knigge}}]{Knigge2012}
{Knigge}, C. 2012, \bibinfo{title}{{Cataclysmic variables in globular clusters.},} \memsai, 83, 549, \dodoi{10.48550/arXiv.1112.1074}

\bibitem[{J.-P. {Lasota}(2001){Lasota}}]{Lasota01}
{Lasota}, J.-P. 2001, \bibinfo{title}{{The disc instability model of dwarf novae and low-mass X-ray binary transients},} \nar, 45, 449, \dodoi{10.1016/S1387-6473(01)00112-9}

\bibitem[{C. Littlefield {et~al.}(2019)Littlefield, Garnavich, Mukai, Mason, Szkody, Kennedy, Myers, \& Schwarz}]{littlefield_fast-cadence_2019}
Littlefield, C., Garnavich, P., Mukai, K., {et~al.} 2019, \bibinfo{title}{Fast-cadence {TESS} {Photometry} and {Doppler} {Tomography} of the {Asynchronous} {Polar} {CD} {Ind}: {A} {Revised} {Accretion} {Geometry} from {Newly} {Proposed} {Spin} and {Orbital} {Periods},} The Astrophysical Journal, 881, 141, \dodoi{10.3847/1538-4357/ab2a17}

\bibitem[{C. {Littlefield} {et~al.}(2023{\natexlab{a}}){Littlefield}, {Mason}, {Garnavich}, {Szkody}, {Thorstensen}, {Scaringi}, {I{\l}kiewicz}, {Kennedy}, \& {Wells}}]{Littlefield23a}
{Littlefield}, C., {Mason}, P.~A., {Garnavich}, P., {et~al.} 2023{\natexlab{a}}, \bibinfo{title}{{SDSS J134441.83+204408.3: A Highly Asynchronous Short-period Magnetic Cataclysmic Variable with a 56 MG Field Strength},} \apjl, 943, L24, \dodoi{10.3847/2041-8213/acaf04}

\bibitem[{C. {Littlefield} {et~al.}(2023{\natexlab{b}}){Littlefield}, {Hoard}, {Garnavich}, {Szkody}, {Mason}, {Scaringi}, {Ilkiewicz}, {Kennedy}, {Rappaport}, \& {Jayaraman}}]{Littlefield23b}
{Littlefield}, C., {Hoard}, D.~W., {Garnavich}, P., {et~al.} 2023{\natexlab{b}}, \bibinfo{title}{{Kepler K2 and TESS Observations of Two Magnetic Cataclysmic Variables: The New Asynchronous Polar SDSS J084617.11+245344.1 and Paloma},} \aj, 165, 43, \dodoi{10.3847/1538-3881/aca1a5}

\bibitem[{N.~R. {Lomb}(1976){Lomb}}]{lomb}
{Lomb}, N.~R. 1976, \bibinfo{title}{{Least-Squares Frequency Analysis of Unequally Spaced Data},} \apss, 39, 447, \dodoi{10.1007/BF00648343}

\bibitem[{G. Myers {et~al.}(2017)Myers, Patterson, Miguel, Hambsch, Monard, Bolt, McCormick, Rea, \& Allen}]{Myers_2017}
Myers, G., Patterson, J., Miguel, E.~d., {et~al.} 2017, \bibinfo{title}{Resynchronization of the Asynchronous Polar CD Ind,} Publications of the Astronomical Society of the Pacific, 129, 044204, \dodoi{10.1088/1538-3873/aa54a8}

\bibitem[{A.~J. Norton {et~al.}(2004)Norton, Wynn, \& Somerscales}]{norton_spin_2004}
Norton, A.~J., Wynn, G.~A., \& Somerscales, R.~V. 2004, \bibinfo{title}{The {Spin} {Periods} and {Magnetic} {Moments} of {White} {Dwarfs} in {Magnetic} {Cataclysmic} {Variables},} The Astrophysical Journal, 614, 349, \dodoi{10.1086/423333}

\bibitem[{J. {Patterson}(1994){Patterson}}]{Patterson94}
{Patterson}, J. 1994, \bibinfo{title}{{The DQ Herculis Stars},} \pasp, 106, 209, \dodoi{10.1086/133375}

\bibitem[{J. {Patterson} {et~al.}(2020){Patterson}, {de Miguel}, {Kemp}, {Dvorak}, {Monard}, {Hambsch}, {Vanmunster}, {Skillman}, {Cejudo}, {Campbell}, {Roberts}, {Jones}, {Cook}, {Bolt}, {Rea}, {Ulowetz}, {Krajci}, {Menzies}, {Lowther}, {Goff}, {Stein}, {Wood}, {Myers}, {Stone}, {Uthas}, {Karamehmetoglu}, {Seargeant}, \& {McCormick}}]{Patterson2020}
{Patterson}, J., {de Miguel}, E., {Kemp}, J., {et~al.} 2020, \bibinfo{title}{{The Spin-period History of Intermediate Polars},} \apj, 897, 70, \dodoi{10.3847/1538-4357/ab863d}

\bibitem[{D. {Pooley} {et~al.}(2002){Pooley}, {Lewin}, {Homer}, {Verbunt}, {Anderson}, {Gaensler}, {Margon}, {Miller}, {Fox}, {Kaspi}, \& {van der Klis}}]{Pooley02}
{Pooley}, D., {Lewin}, W. H.~G., {Homer}, L., {et~al.} 2002, \bibinfo{title}{{Optical Identification of Multiple Faint X-Ray Sources in the Globular Cluster NGC 6752: Evidence for Numerous Cataclysmic Variables},} \apj, 569, 405, \dodoi{10.1086/339210}

\bibitem[{N. {Rea} {et~al.}(2017){Rea}, {Coti Zelati}, {Esposito}, {D'Avanzo}, {de Martino}, {Israel}, {Torres}, {Campana}, {Belloni}, {Papitto}, {Masetti}, {Carrasco}, {Possenti}, {Wieringa}, {Wilhelmi}, {Li}, {Bozzo}, {Ferrigno}, {Linares}, {Tauris}, {Hernanz}, {Ribas}, {Monelli}, {Borghese}, {Baglio}, \& {Casares}}]{Rea17}
{Rea}, N., {Coti Zelati}, F., {Esposito}, P., {et~al.} 2017, \bibinfo{title}{{Multiband study of RX J0838-2827 and XMM J083850.4-282759: a new asynchronous magnetic cataclysmic variable and a candidate transitional millisecond pulsar},} \mnras, 471, 2902, \dodoi{10.1093/mnras/stx1560}

\bibitem[{L.~E. {Rivera Sandoval} {et~al.}(2018){Rivera Sandoval}, {van den Berg}, {Heinke}, {Cohn}, {Lugger}, {Anderson}, {Cool}, {Edmonds}, {Wijnands}, {Ivanova}, \& {Grindlay}}]{RiveraSandoval18}
{Rivera Sandoval}, L.~E., {van den Berg}, M., {Heinke}, C.~O., {et~al.} 2018, \bibinfo{title}{{New cataclysmic variables and other exotic binaries in the globular cluster 47 Tucanae$^{*}$},} \mnras, 475, 4841, \dodoi{10.1093/mnras/sty058}

\bibitem[{J.~D. {Scargle}(1982){Scargle}}]{scargle}
{Scargle}, J.~D. 1982, \bibinfo{title}{{Studies in astronomical time series analysis. II. Statistical aspects of spectral analysis of unevenly spaced data.},} \apj, 263, 835, \dodoi{10.1086/160554}

\bibitem[{R. {Schwarz} {et~al.}(2007){Schwarz}, {Schwope}, {Staude}, {Rau}, {Hasinger}, {Urrutia}, \& {Motch}}]{Schwarz07}
{Schwarz}, R., {Schwope}, A.~D., {Staude}, A., {et~al.} 2007, \bibinfo{title}{{Paloma (RX J0524+42): the missing link in magnetic CV evolution?},} \aap, 473, 511, \dodoi{10.1051/0004-6361:20077684}

\bibitem[{A.~D. {Schwope} {et~al.}(2024){Schwope}, {Knauff}, {Kurpas}, {Salvato}, {Stelzer}, {St{\"u}tz}, \& {Tub{\'\i}n-Arenas}}]{Schwope24}
{Schwope}, A.~D., {Knauff}, K., {Kurpas}, J., {et~al.} 2024, \bibinfo{title}{{A first systematic characterization of cataclysmic variables in SRG/eROSITA surveys},} \aap, 690, A243, \dodoi{10.1051/0004-6361/202450537}

\bibitem[{M.~M. {Shara} {et~al.}(1996){Shara}, {Bergeron}, {Gilliland}, {Saha}, \& {Petro}}]{Shara96}
{Shara}, M.~M., {Bergeron}, L.~E., {Gilliland}, R.~L., {Saha}, A., \& {Petro}, L. 1996, \bibinfo{title}{{Cataclysmic and Close Binaries in Star Clusters. IV. The Unexpectedly Low Number of Erupting Dwarf Novae Detected by the Hubble Space Telescope in the Core of 47 Tucanae},} \apj, 471, 804, \dodoi{10.1086/178008}

\bibitem[{V.~F. {Suleimanov} {et~al.}(2019){Suleimanov}, {Doroshenko}, \& {Werner}}]{Suleimanov19}
{Suleimanov}, V.~F., {Doroshenko}, V., \& {Werner}, K. 2019, \bibinfo{title}{{Hard X-ray view on intermediate polars in the Gaia era},} \mnras, 482, 3622, \dodoi{10.1093/mnras/sty2952}

\bibitem[{G. {Tovmassian} {et~al.}(2017){Tovmassian}, {Gonz{\'a}lez-Buitrago}, {Thorstensen}, {Kotze}, {Breytenbach}, {Schwope}, {Bernardini}, {Zharikov}, {Hernandez}, {Buckley}, {de Miguel}, {Hambsch}, {Myers}, {Goff}, {Cejudo}, {Starkey}, {Campbell}, {Ulowetz}, {Stein}, {Nelson}, {Reichart}, {Haislip}, {Ivarsen}, {LaCluyze}, {Moore}, \& {Miroshnichenko}}]{Tovmassian17}
{Tovmassian}, G., {Gonz{\'a}lez-Buitrago}, D., {Thorstensen}, J., {et~al.} 2017, \bibinfo{title}{{IGR J19552+0044: A new asynchronous short period polar. Filling the gap between intermediate and ordinary polars},} \aap, 608, A36, \dodoi{10.1051/0004-6361/201731323}

\bibitem[{F. {Verbunt} \& G. {Hasinger}(1998){Verbunt} \& {Hasinger}}]{Verbunt98}
{Verbunt}, F., \& {Hasinger}, G. 1998, \bibinfo{title}{{Nine X-ray sources in the globular cluster 47 Tucanae},} \aap, 336, 895, \dodoi{10.48550/arXiv.astro-ph/9805348}

\bibitem[{P. Virtanen {et~al.}(2020)Virtanen, Gommers, Oliphant, Haberland, Reddy, Cournapeau, Burovski, Peterson, Weckesser, Bright, {van der Walt}, Brett, Wilson, Millman, Mayorov, Nelson, Jones, Kern, Larson, Carey, Polat, Feng, Moore, {VanderPlas}, Laxalde, Perktold, Cimrman, Henriksen, Quintero, Harris, Archibald, Ribeiro, Pedregosa, {van Mulbregt}, \& {SciPy 1.0 Contributors}}]{2020SciPy-NMeth}
Virtanen, P., Gommers, R., Oliphant, T.~E., {et~al.} 2020, \bibinfo{title}{{{SciPy} 1.0: Fundamental Algorithms for Scientific Computing in Python},} Nature Methods, 17, 261, \dodoi{10.1038/s41592-019-0686-2}

\bibitem[{Q. Wang {et~al.}(2020)Wang, Qian, Han, Fang, Zang, \& Liu}]{Wang_2020}
Wang, Q., Qian, S., Han, Z., {et~al.} 2020, \bibinfo{title}{Spot Model for Identifications of Periods in Asynchronous Polars,} The Astrophysical Journal, 892, 38, \dodoi{10.3847/1538-4357/ab7759}

\bibitem[{B. {Warner}(1995){Warner}}]{Warner95}
{Warner}, B. 1995, {Cataclysmic variable stars}, Vol.~28

\bibitem[{M.~C. {Weisskopf} {et~al.}(2002){Weisskopf}, {Brinkman}, {Canizares}, {Garmire}, {Murray}, \& {Van Speybroeck}}]{Weisskopf02}
{Weisskopf}, M.~C., {Brinkman}, B., {Canizares}, C., {et~al.} 2002, \bibinfo{title}{{An Overview of the Performance and Scientific Results from the Chandra X-Ray Observatory},} \pasp, 114, 1, \dodoi{10.1086/338108}

\bibitem[{G.~A. {Wynn} \& A.~R. {King}(1992){Wynn} \& {King}}]{wynn}
{Wynn}, G.~A., \& {King}, A.~R. 1992, \bibinfo{title}{{Theoretical X-ray power spectra of intermediate polars},} \mnras, 255, 83, \dodoi{10.1093/mnras/255.1.83}

\end{thebibliography}
\bibliographystyle{aasjournalv7}



\end{document}